\documentclass[11pt]{article}
\usepackage{amssymb}
\usepackage{amsthm}
\usepackage{graphicx}

\newlength{\extraspace}
\newlength{\extraspaces}
\newcommand{\be}{\begin{equation}
\addtolength{\abovedisplayskip}{\extraspaces}
\addtolength{\belowdisplayskip}{\extraspaces}
\addtolength{\abovedisplayshortskip}{\extraspace}
\addtolength{\belowdisplayshortskip}{\extraspace}}
\newcommand{\ee}{\end{equation}}

\newcommand{\ba}{\begin{eqnarray}
\addtolength{\abovedisplayskip}{\extraspaces}
\addtolength{\belowdisplayskip}{\extraspaces}
\addtolength{\abovedisplayshortskip}{\extraspace}
\addtolength{\belowdisplayshortskip}{\extraspace}}
\newcommand{\ea}{\end{eqnarray}}

\newcounter{saveeqn}

\begin{document}
\addtolength{\baselineskip}{1.5mm}

\thispagestyle{empty}

\vbox{}
\vspace{1.6cm}

\begin{center}
{\LARGE{A Proof for a Theorem of Wald in Arbitrary Dimensions}}
\vspace{2mm}

\vspace{16mm}
{H. S. Tan }
\\[6mm]
{\it Berkeley Center for Theoretical Physics and Department of Physics,\\[1mm]
University of California, Berkeley,
CA 94720-7300, U.S.A.}\\[15mm]

\end{center}
\vspace{0.5cm}

\centerline{\bf Abstract}\bigskip
\noindent
Static, axisymmetric solutions form a large class of important black holes in classical GR. In four dimensions, the existence of their most general metric ansatz relies on the fact that two-dimensional subspaces of the tangent space at each point spanned by vectors orthogonal to the time-translation and rotation Killing fields are integrable. This was first proved by Wald via an application of Frobenius theorem. In this note, we furnish an elementary proof for this theorem by Wald in arbitrary dimensions which yields the metric ansatz for the most general solution of the $D-$dimensional vacuum Einstein equations that admits $D-2$ orthogonal and commuting Killing vector fields.

\newpage
\section{Motivations}
We begin our discussion in the context of four-dimensional pure gravity in vacuum. In 1917, Hermann Weyl
showed that any static, axisymmetric space-time satisfying Einstein's vacuum field equations $R_{\mu \nu} = 0$
can be expressed in the form \cite{{Weyl},{Islam}}
\be
{\rm d}s^2 = -{\rm e}^{2U} {\rm d}t^2 + {\rm e}^{-2U} \bigg( {\rm e}^{2\gamma}({\rm d}r^2+{\rm d}z^2) + r^2 {\rm d}
\phi^{2} \bigg)\,,
\ee
where $U(r,z)$ is an arbitrary axisymmetric solution of Laplace equation in 3D flat space
\be
\label{Laplace}
\frac{\partial^2 U}{\partial r^2} + \frac{1}{r}\frac{\partial U}{\partial r} + \frac{\partial^2 U}{\partial z^2} = 0\,,
\ee
and $\gamma(r,z)$ satisfies
\be
\label{quadratures}
\frac{\partial \gamma}{\partial r} = r \Bigg[ \bigg( \frac{\partial U}{\partial r} \bigg)^2 - \bigg( \frac{\partial U}{\partial z} \bigg)^2 \Bigg]\,,
\qquad \frac{\partial \gamma}{\partial z} = 2r\frac{\partial U}{\partial r} \frac{\partial U}{\partial z} \,,
\ee
the solution of which is given by a line integral. This is the first general family of exact solutions to the field
equations which are notoriously non-linear and heavily coupled \cite{{Kramer},{Wald}}. Since $U$ is harmonic, it can be regarded
as a Newtonian potential produced by axisymmetric sources. In this manner, they have a simple and beautiful
relationship to the corresponding Newtonian solutions: each Weyl solution arises from a Newtonian static
and axisymmetric vacuum potential. Perhaps the mathematically simplest example is the Curzon-Chazy solution \cite{{Bonnor1},{Bicak}} in which $U=-\frac{m}{\sqrt{r^2+z^2}}$ is the Newtonian potential of a spherical point particle. The space-time however is not of spherical symmetry, and indeed, in general, we must exercise caution when interpreting these Weyl metrics --- for there is no correspondence between the geometry of the physical source and the geometry of the Newtonian source \cite{{Bonnor1},{Hillman}}. Other 4D examples of Weyl solutions include the well-known Schwarzschild solution \cite{Islam} and the C-metric which describes two black holes accelerating apart \cite{Kinnersley}.

Since the Weyl class contains many physically interesting solutions, it is natural to seek its higher-dimensional
analogue. There are several ways to do this. One possibility is to seek the class of $D-$dimensional solutions that
are static and axisymmetric, in the sense that they admit an isometry group $\Re \times O(D-2)$, but this has
been attempted without success in \cite{Myers1}. Alternatively, we observe that Weyl solutions can be characterized
as having two orthogonal commuting Killing vector fields. Hence, a way to generalize Weyl class to higher
dimensions would be to seek all solutions of the vacuum Einstein equations that admit $D-2$ orthogonal, commuting
Killing vector fields. In 2001, this was achieved by Roberto Emparan and Harvey Reall in \cite{Emparan1}. It was found
that as in four dimensions, the higher-dimensional Weyl class of solutions can also be reduced to the axisymmetric
Laplace equation in 3D flat space and a pair of quadratures.

To be explicit, the generalized Weyl class of solutions in $D$ dimensions is parametrized by $D-3$
harmonic functions in an auxiliary flat space. These harmonic functions can be interpreted as Newtonian
potentials of axisymmetric sources. Now, there are many solutions to the 3D Laplace equation \cite{Hillman}.
To select those that might be of physical importance, one recalls that in four dimensions, interesting
Weyl solutions such as the Schwarzschild solution, have sources of the same form --- rods of vanishing thickness
 on the axis of symmetry. Deviations from this form often lead to solutions that are nakedly singular.
An example is the previously mentioned Curzon-Chazy solution which was shown in \cite{Gautreau} to exhibit a directional singularity. It turns out that in the generalized approach, the harmonic functions of known $D>4$ Weyl solutions have their harmonic functions corresponding also to thin rods on the axis of symmetry in the auxiliary 3D flat space - this naturally
holds true for $4D$ Weyl solutions although this constraint had never been identified prior to \cite{Emparan1}. These `generalized' Weyl solutions include black rings and black saturns.

To generalize $4D$ Weyl solutions to arbitrary dimensions, it is fundamental to work in a convenient coordinate
chart for the general $D$-dimensional line element admitting $D-2$ commuting Killing vector fields. In four dimensions
this was done by Wald in \cite{Wald} via a beautiful application of Frobenius theorem. The aim of this note is to generalize the 4D theorem in \cite{Wald} to arbitrary dimensions and show that the line element of generalized Weyl solutions can be written in a simple form that has enabled Emparan and Reall to extract Laplace equation from Einstein's field equations quite directly. This generalized theorem has been stated in \cite{Emparan1} but to the best of our knowledge, no explicit proof has ever been constructed \footnote{The proof was considered in \cite{Emparan1} to be a straightforward generalization of that in \cite{Wald}.}.
\section{The Proof}
Assume that the metric is Riemannian or Lorentzian and let $\xi_{(i)}$ denote the Killing vector fields
$1\leq i \leq D-2$. Since these vectors commute, it is possible to choose coordinates $(x^{i},y^{1},y^{2})$
such that $\xi_{(i)} = \frac{\partial}{\partial x^{i}}$ with the metric coefficients as functions of $y^{1}$ and $y^{2}$.
Then, one proceed to show that the coordinates $y^{1}$ and $y^{2}$ can be chosen to span two-dimensional
surfaces orthogonal to all the Killing vector fields ( i.e. there are no cross-terms $dx^{i} dy^{1}$ and
$dx^{i} dy^{2}$ in the line element. ) To do this, one has to show that the two-dimensional subspaces
of the tangent space orthogonal to all $\xi_{i}$ are integrable. Sufficient conditions for integrability are supplied
by the following theorem:

\newtheorem*{theorem}{Theorem 1}
\begin{theorem}
\textrm{Let $\xi_{(i)}, 1 \leq i \leq D-2$ be commuting Killing vector fields such that for each $i$,\\
$(a)  \xi_{(1)}^{[\mu_{1}} \xi_{(2)}^{\mu_2} \dots \xi_{(D-2)}^{\mu_{D-2}} \nabla^{v} \xi_{(i)}^{p]}$ vanishes at at least one point,\\
$(b)  \xi_{(i)}^{v}R_{v}^{[p}\xi_{(1)}^{\mu_{1}}\xi_{(2)}^{\mu_{2}} \dots \xi_{(D-2)}^{\mu_{(D-2)}]} = 0$.\\
Then, the two-dimensional planes orthogonal to all $\xi_{(i)}$ are integrable.}
\end{theorem}

This theorem was first stated in \cite{Emparan1} but no proof was given. In this note, we present
its first explicit proof. To prove Theorem 1, we begin by invoking Frobenius theorem in the form of cotangent vector fields, which states \cite{Wald}

\newtheorem*{theorem2}{Frobenius Theorem}
\begin{theorem2}
\textrm{ Let $T^{*}$ be a smooth specification of an $(n-m)-$dimensional subspace of 1-forms. Then, the associated
$m$-dimensional subspace $S$ of the tangent space admits integrable submanifolds if and only if $\forall \mathbf{Y} \in T^{*}$,
we have $d\mathbf{Y} = \sum_{\alpha} \mathbf{U}^{\alpha} \wedge \mathbf{V}^{\alpha}$ or $\nabla_{[a}\mathbf{Y}_{b]} = \sum_{\alpha =1}^{n-m} \mathbf{U}_{[a}^{\alpha} \mathbf{V}_{b]}^{\alpha}$
where each $\mathbf{U}^{\alpha}$ is an arbitrary one-form and each $\mathbf{V}^{\alpha} \in T^{*}$.}
\end{theorem2}
We proceed by letting $\xi_{(i)} = Y $ in Frobenius theorem, hence
\be
\nabla_{[a} \xi_{(i)b]} = \sum_{\alpha =1}^{D-2} U_{(i)[a}^{\alpha} \xi_{b]}^{\alpha} \quad \forall 1\leq i \leq D-2\,.
\ee
This is equivalent to the condition that
\be
\xi_{(1)}^{[\mu_1}\,\xi_{(2)}^{\mu_2} \dots \xi_{(i-1)}^{\mu_{i-1}}\,\xi_{(i+1)}^{\mu_{i+1}} \dots \xi_{(D-2)}^{\mu_{D-2}}\,\xi_{(i)}^{\mu_i}\,\nabla_{v}\xi_{(i)}^{p]} = 0 \,\, \forall 1\leq i \leq D-2\,,
\ee
which is in turn equivalent to
\be
\epsilon_{\mu_1 \mu_2 \dots \mu_{i-1} \mu_{i+1} \dots \mu_{D-2} \mu_i vp} \xi_{(1)}^{\mu_1}\,\xi_{(2)}^{\mu_2} \dots \xi_{(i-1)}^{\mu_{i-1}}\,\xi_{(i+1)}^{\mu_{i+1}} \dots \xi_{(D-2)}^{\mu_{D-2}}\,\xi_{(i)}^{\mu_i}\,\nabla_{v}\xi_{(i)}^{p} = 0\,,
\ee
where $\epsilon_{\mu_1 \mu_2 \dots \mu_{i-1} \mu_{i+1} \dots \mu_{D-2} \mu_i vp}$ is the completely antisymmetric
volume-element of the $D-$dimensional metric. Define a generalized twist tensor $W_{(i)}$ of $\xi_{(i)}$ by
\be
W_{(i)\mu_1 \mu_2 \dots \mu_{i-1} \mu_{i+1} \dots \mu_{D-2}} = \epsilon_{\mu_1 \mu_2 \dots \mu_{i-1} \mu_{i+1} \dots \mu_{D-2} \mu_i vp} \xi_{(i)}^{\mu_i} \nabla^{v} \xi_{(i)}^{p}\,.
\ee
Then (2.3) can be written as
\be
\xi_{(1)}^{\mu_1}\,\xi_{(2)}^{\mu_2} \dots \xi_{(i-1)}^{\mu_{i-1}}\,\xi_{(i+1)}^{\mu_{i+1}} \dots \xi_{(D-2)}^{\mu_{D-2}} W_{(i)\mu_1 \mu_2 \dots \mu_{i-1} \mu_{i+1} \dots \mu_{D-2}} = 0\,\, \forall i\,.
\ee
By our hypothesis in Theorem 1, the left hand side of (2.5) vanishes at at least one point of the manifold. It will
vanish everywhere if its covariant derivative is identically zero. Taking the covariant derivative, we have
\ba
&&\nabla_{h} \bigg( \xi_{(1)}^{\mu_1}\,\xi_{(2)}^{\mu_2} \dots \xi_{(i-1)}^{\mu_{i-1}}\,\xi_{(i+1)}^{\mu_{i+1}} \dots \xi_{(D-2)}^{\mu_{D-2}} W_{(i)\mu_1 \mu_2 \dots \mu_{i-1} \mu_{i+1} \dots \mu_{D-2} \mu_i vp} \bigg)
\cr
&=&\bigg( \xi_{(1)}^{\mu_1}\,\xi_{(2)}^{\mu_2} \dots \xi_{(i-1)}^{\mu_{i-1}}\,\xi_{(i+1)}^{\mu_{i+1}} \dots \xi_{(D-2)}^{\mu_{D-2}} \bigg) \nabla_{h} W_{(i)\mu_1 \mu_2 \dots \mu_{i-1} \mu_{i+1} \dots \mu_{D-2}}
\cr
&&+ W_{(i)\mu_1 \mu_2 \dots \mu_{i-1} \mu_{i+1} \dots \mu_{D-2}} \nabla_{h} \bigg( \xi_{(1)}^{\mu_1}\,\xi_{(2)}^{\mu_2} \dots \xi_{(i-1)}^{\mu_{i-1}}\,\xi_{(i+1)}^{\mu_{i+1}} \dots \xi_{(D-2)}^{\mu_{D-2}} \bigg)
\ea
by the linearity of the covariant derivative. One will effectively prove Theorem 1 if (2.6) vanishes.
Now, recall that the Lie derivative of a $(0,q)$ tensor $T$ along vector $V^{\mu}$ is the
sum of two kinds of terms : the directional derivative of $T$ along $V$ and $q-$terms involving the
covariant derivative of $V$ contracted with each of the lower indices. Explicitly, one has
\ba
\pounds_{V} T_{a_1a_2 \dots a_q} &=& V^{\sigma} \nabla_{\sigma} T_{a_1a_2 \dots a_q}
\cr
&&+ \underbrace{T_{\hat{a} a_2 \dots a_q} \nabla_{a_1} V^{\hat{a}} + T_{a_1 \hat{a} a_3 \dots a_q} \nabla_{a_2} V^{\hat{a}} + \dots + T_{a_1 a_2 \dots a_{q-1}\hat{a}} \nabla_{a_q} V^{\hat{a}}}_{\textrm{$q$ terms}}
\ea
From (2.6) and (2.7), it is just a straightforward though tedious rearrangement of terms to obtain more suggestively
\ba
&&\nabla_{h} \bigg( \xi_{(1)}^{\mu_1}\,\xi_{(2)}^{\mu_2} \dots \xi_{(i-1)}^{\mu_{i-1}}\,\xi_{(i+1)}^{\mu_{i+1}} \dots \xi_{(D-2)}^{\mu_{D-2}} W_{(i)\mu_1 \mu_2 \dots \mu_{i-1} \mu_{i+1} \dots \mu_{D-2} \mu_i vp} \bigg)
\cr
&=& (D-3) \bigg( \xi_{[(1)}^{\mu_1} \dots \xi_{(l-1)}^{\mu_{l-1}} \xi_{(l+1)}^{\mu_{l+1}}\dots \xi_{(i-1)}^{\mu_{i-1}}\,\xi_{(i+1)}^{\mu_{i+1}} \dots \xi_{(D-2)}^{\mu_{D-2}}\pounds_{\xi_l ]} W_{(i)\mu_1 \dots \mu_{l-1} \mu_{l+1}\dots \mu_{i-1} \mu_{i+1} \dots \mu_{D-2} h} \bigg)
\cr
&&+(D-3)! \bigg( W_{fh \mu_2 \dots \mu_{i-1} \mu_{i+1}\dots \mu_{D-2} } \xi_{[(1)}^{\alpha} \xi_{(2)}^{\mu_2} \dots \xi_{(i-1)}^{\mu_{i-1}} \xi_{(i+1)}^{\mu_{i+1}} \dots \nabla_{\alpha} \xi_{(D-2)]}^{f} \bigg)
\cr
&&+(D-2) \bigg( \xi_{(1)}^{\mu_1} \dots \xi_{(i-1)}^{\mu_{i-1}} \xi_{(i+1)}^{\mu_{i+1}} \dots \xi_{(D-2)}^{\mu_{(D-2)}} \nabla_{[h} W_{\mu_1 \dots \mu_{i-1} \mu_{i+1} \dots \mu_{D-2}]} \bigg)
\ea
Now, the first term after the equality sign in (2.8) naturally vanishes. To see this, one recalls that $W$ is
a tensor field constructed from the tensors $\xi_{(i)}$ and $g_{\mu \nu}$ with $i\neq l$.
The group of diffeomorphisms generated by $\xi_{(l)}$ leaves $\xi_{(i)}$ invariant since both commute
with each other. It also leaves the metric invariant by definition since $\xi_{(l)}$ is a Killing vector. Hence this
group of diffeomorphisms leaves invariant any tensor field constructed just out of $\xi_{(i)}$ and the metric.
Therefore,
\be
\pounds_{\xi_{(l)}} W_{(i)\mu_1 \dots \mu_{l-1}\mu_{l+1} \dots \mu_{i-1} \mu_{i+1} \dots \mu_{D-2} h} = 0
\ee
which leads to the first term after the equality sign in (2.8) to be zero. The second term vanishes easily
since for two commuting vector fields $\xi_{(l)}$ and $\xi_{(k)}$, we have
\be
2\xi_{[(l)}^{\sigma} \nabla_{\sigma} \xi_{(k)]}^{\mu} = \xi_{(l)}^{\sigma} \nabla_{\sigma} \xi_{(k)}^{\mu} - \xi_{(k)}^{\sigma} \nabla_{\sigma} \xi_{(l)}^{\mu} = \pounds_{\xi_{(l)}} \xi_{(k)}^{\mu} = 0
\ee
since the Lie derivative along one vector of another vector is simply their commutation. However, that the
third term disappears as well is not so immediate. First, we recall that for a general $n$-dimensional Riemannian
metric with $s$ minuses appearing in the signature of $g_{\mu \nu}$, we have the relation for the volume element
\be
\epsilon^{a_1 \dots a_n} \epsilon_{b_1 \dots b_n} = (-1)^{s} n! \delta^{[a_1}_{b_1} \delta^{a_2}_{b_2} \dots \delta^{a_n ]}_{b_n}
\ee
from which contraction over $j$ of its indices yields
\be
\epsilon^{a_1 \dots a_j a_{j+1} \dots a_n} \epsilon_{a_1 \dots a_j b_{j+1} \dots b_n} = (-1)^{s} (n-j)!j!\delta^{[a_{j+1}}_{b_{j+1}} \dots \delta^{a_n]}_{b_n}
\ee
from which we can write
\ba
&&\nabla_{[h} W_{\mu_1 \dots \mu_{i-1} \mu_{i+1} \dots \mu_{D-2}]}
\cr
&=&\delta^{ [ \hat{h}}_{h} \delta^{\hat{\mu}_1}_{\mu_1} \delta^{\hat{\mu}_2}_{\mu_2} \dots \delta^{\hat{\mu}_{i-1}}_{\mu_{i-1}} \delta^{\hat{\mu}_{i+1}}_{\mu_{i+1}} \dots \delta^{\hat{\mu}_{D-2} ] }_{\mu_{D-2}} \nabla_{\hat{h}}W_{\hat{\mu}_1 \dots \hat{\mu}_{i-1} \hat{\mu}_{i+1} \dots \hat{\mu}_{D-2}}
\cr
&=&-\frac{1}{2(D-2)!} \epsilon^{ab\hat{h}\hat{\mu}_1 \dots \hat{\mu}_{i-1}\hat{\mu}_{i+1} \dots \hat{\mu}_{D-2}} \epsilon_{abh\mu_1 \dots \mu_{i-1}\mu_{i+1} \dots \mu_{D-2}} \nabla_{\hat{h}} W_{\hat{\mu}_1 \dots \hat{\mu}_{i-1} \hat{\mu}_{i+1} \dots \hat{\mu}_{D-2}}
\ea
Then, we consider
\ba
&&\epsilon^{ab\hat{h}\hat{\mu}_1 \dots \hat{\mu}_{i-1}\hat{\mu}_{i+1} \dots \hat{\mu}_{D-2}}  \nabla_{\hat{h}} W_{\hat{\mu}_1 \dots \hat{\mu}_{i-1} \hat{\mu}_{i+1} \dots \hat{\mu}_{D-2}}
\cr
&=&\epsilon^{\hat{\mu}_1 \dots \hat{\mu}_{i-1}\hat{\mu}_{i+1} \dots \hat{\mu}_{D-2}ab\hat{h}} \epsilon_{\hat{\mu}_1 \dots \hat{\mu}_{i-1}\hat{\mu}_{i+1} \dots \hat{\mu}_{D-2}\mu_i vp} \nabla_{\hat{h}} \bigg( \xi_{(i)}^{\mu_{i}} \nabla^{v} \xi^{p}_{(i)} \bigg)
\cr
&=&-6 (D-3)! \delta^{[a}_{\mu_{i}} \delta^{b}_{v} \delta^{\hat{h}]}_{p} \nabla_{\hat{h}}\bigg( \xi_{(i)}^{\mu_{i}} \nabla^{v} \xi_{(i)}^{p} \bigg)
\cr
&=&-2(D-3)! \nabla_{\hat{h}} \bigg( \xi_{(i)}^{a} \nabla^{b} \xi_{(i)}^{\hat{h}} + \xi_{(i)}^{b} \nabla^{\hat{h}} \xi_{(i)}^{a} + \xi_{(i)}^{\hat{h}} \nabla^{a} \xi_{(i)}^{b} \bigg)\,.
\ea
To simplify the above expression, we make use of the following identity for Killing vector fields
\be
\nabla_{a} \nabla_{b} \xi_{c} = - {R_{bca}}^{d} \xi_{d}\,.
\ee
Contracting (2.15) over $a$ and $b$ yields
\be
\nabla^{a} \nabla_{a} \xi_{c} = -{R_{c}}^{d} \xi_{d}\,.
\ee
With (2.15), the final line of (2.14) becomes
\ba
&&-2(D-3)! \bigg( -\big(\nabla^{\hat{h}} \xi^{a}_{(i)} \big) \big( \nabla^{\hat{h}} \xi^{b}_{(i)} \big) + \xi^{a}_{(i)} \nabla_{\hat{h}} \nabla^{b} \xi^{\hat{h}}_{(i)} + \xi^{b}_{(i)} \nabla_{\hat{h}} \nabla^{\hat{h}} \xi^{a}_{(i)} - \xi^{\hat{h}}_{(i)} R^{ab}_{\hat{h}d}  \xi^{d}_{(i)} \bigg)
\cr
&=& -2(D-3)! \left( -\xi^{a}_{(i)} \nabla_{\hat{h}} \nabla^{\hat{h}} \xi^{b}_{(i)}  -\xi^{b}_{(i)} \nabla_{\hat{h}} \nabla^{\hat{h}} \xi^{a}_{(i)} \right) \,\,\,\, (\textrm{since ${R^{ab}}_{\hat{h}d} = -{R^{ab}}_{d \hat{h}}$})
\cr
&=& 4(D-3)! \left( \xi_{(i)}^{[b} R^{a]}_{\hat{h}} \xi^{\hat{h}}_{(i)} \right) \,\,\,\, (\textrm{by} 2.16)
\ea
Hence, we have obtained from (2.13)-(2.17)
\be
\nabla_{[h} W_{\mu_1 \dots \mu_{i-1} \mu_{i+1} \dots \mu_{D-2}]} = \epsilon_{abh\mu_1 \dots \mu_{i-1}\mu_{i+1} \dots \mu_{D-2}} \frac{2(D-3)!}{(D-2)!} \xi^{[b}_{(i)} R^{a]}_{c} \xi^{c}_{(i)}\,.
\ee
Finally, we find
\ba
&&\nabla_{h} \bigg( \xi_{(1)}^{\mu_1} \dots \xi_{(i-1)}^{\mu_{i-1}} \xi_{(i+1)}^{\mu_{i+1}} \dots \xi_{(D-2)}^{\mu_{(D-2)}} \nabla_{[h} W_{\mu_1 \dots \mu_{i-1} \mu_{i+1} \dots \mu_{D-2}]} \bigg)
\cr
&=& 2\epsilon_{abh\mu_1 \dots \mu_{i-1}\mu_{i+1} \dots \mu_{D-2}} \xi_{(1)}^{\mu_1} \dots \xi_{(i-1)}^{\mu_{i-1}} \xi_{(i+1)}^{\mu_{i+1}} \dots \xi_{(D-2)}^{\mu_{(D-2)}} \xi^{[b}_{(i)} R^{a]}_{c} \xi^{c}_{(i)} = 0
\ea
by the hypothesis stated earlier in Theorem 1. Thus we have shown that all three terms of (2.8) vanish. This completes
an explicit proof of Theorem 1. The reader can compare with the four-dimensional case in \cite{Wald}, and see that what we have done is quite a straightforward generalization, as was pointed out in \cite{Emparan1}.

We have, in the beginning, wanted to show the coordinates $y^{1}$ and $y^{2}$ can be chosen to span two-dimensional
surfaces orthogonal to all the Killing vector fields. Following the arguments presented in \cite{Emparan1}, let us observe if the generalized Weyl class satisfies the two sufficient conditions for integrability as specified by Theorem 1. Since we are considering vacuum
solutions of the Einstein equations, $R_{\mu \nu} = 0$ and condition (b) of Theorem 1 is satisfied trivially.
Condition (a) is less obvious. In four dimensions, it is usually assumed that one of the Killing vector fields is
an angle corresponding to rotations about an axis of symmetry, and thus it vanishes on this axis which is
invariant under this rotation. The same assumption can be used to satisfy condition (a) in higher dimensions.

Since conditions of Theorem 1 are satisfied, then the coordinates $y^{1}$ and $y^{2}$ can be chosen to
span the two-dimensional surfaces orthogonal to the Killing vector fields. If it is further assumed that
the commuting Killing vector fields are orthogonal to each other (which defines the generalized Weyl class)
then the metric takes the simple form :
\be
{\rm d}s^2 = \sum_{i=1}^{D-2} \varepsilon_i {\rm e}^{2U_i} ({\rm d}x^{i})^2 + g_{ab} {\rm d}y^{a} {\rm d}y^{b}\,,
\ee
where $a$ and $b$ take the values $1,2$, the metric coefficients are all independent of $x^{i}$ and
$\varepsilon_{i} = 1$ or $ -1$ depending on whether $\xi_{i}$ is space-like or time-like respectively. Finally, we can choose
coordinates such that
\be
g_{ab} {\rm d}y^{a} {\rm d}y^{b} = e^{2C} {\rm d}Z {\rm d} \bar{Z}\,,
\ee
where $Z$ and $\bar{Z}$ are complex conjugate co-ordinates if the transverse space is space-like --- which Emparan and Reall assume in their seminal work \cite{Emparan1}. Indeed, it is always possible to find coordinates such that we have (2.21).
This is because it can be easily shown that any two-dimensional Riemannian manifold is conformally flat
(see for example \cite{Akivis}).

We have thus arrived at a simple form for a $D-$dimensional metric that has
$D-2$ orthogonal commuting Killing vector fields. In \cite{Emparan1}, this was used as the crucial starting point for constructing generalized Weyl solutions. There are no cross terms between the Killing and non-Killing differentials
in the metric because as we have proven --- the 2-surfaces can be chosen to be orthogonal to all the orbits of
the Killing vector fields. This is not a trivial result that carries over to, for example, a metric of $D-3$ commuting
and orthogonal Killing vectors. Notice from (2.2) or (2.3) that $D-2$ is `just enough' for us to use the volume
element which has to have $D$ indices as a completely antisymmetric tensor.

\bigskip\bigskip\centerline{{\bf Acknowledgment}}
\nobreak\noindent The author is grateful to Prof. Ori Ganor and Prof. Edward Teo for kind encouragements.
{\renewcommand{\Large}{\normalsize}


\begin{thebibliography}{99}

\bibitem{Weyl}
H.~Weyl,
Ann.\ Phys.(Leipzig)\ B {\bf 54} (1917) 117.

\bibitem{Islam}
J.~N.~Islam,
``Rotating fields in general relativity,''
(Cambridge University Press, Cambridge, 1985).

\bibitem{Kramer}
D.~Kramer, H.~Stephani, M.~MacCallum and E.~Herlt,
``Exact solutions of Einstein's field equations,''
(Cambridge University Press, Cambridge, 1980).

\bibitem{Wald}
R.~M~Wald,
``General Relativity,''
(University of Chicago Press, USA, 1984).

\bibitem{Bonnor1}
W.~B.~Bonnor,
``Physical interpretation of vacuum solutions of Einstein's equations. Part 1: Time-independent solutions,''
Gen.\ Rel.\ Grav. {\bf 24} (1992) 551.

\bibitem{Bicak}
J.~Bicak,
``Selected solutions of Einstein's field equations: their role in general relativity and astrophysics,"
(Springer-Verlag, Berlin-New York, 2000).

\bibitem{Hillman}
C.~Hillman,
``A collection of expository postings on 4D Weyl solutions,''
(http://math.ucr.edu/home/baez/PUB/weylvac).

\bibitem{Kinnersley}
W.~Kinnersley and M.~Walker,
``Uniformly accelerating charged mass in general relativity,''
Phys.\ Rev.\ D {\bf 2} (1970) 1359

\bibitem{Myers1}
R.~C.~Myers,
``Higher-dimensional black holes in compactified space-times,''
Phys.\ Rev.\ D {\bf 35} (1987) 455

\bibitem{Emparan1}
R.~Emparan and H.~S.~Reall,
``Generalized Weyl solutions,"
Phys. \ Rev. \ D {\bf 65} (2002) 084025
[arXiv:hep-th/0110258]
%%CITATION = HEP-TH 0110258

\bibitem{Gautreau}
R.~Gautreau and J.~L.~Anderson,
Phys. \ Lett. {\bf 25A} (1967) 291


\bibitem{Akivis}
M.~A.~Akivis and V.~V.~Goldberg
``Conformal differential geometry and its generalizations,"
(New York, Wiley, 1996).

\end{thebibliography}
\end{document}